# Epitaxial Graphene Nanoribbons on Bunched Steps of a 6H-SiC(0001) Substrate: Aromatic Ring Pattern and Van Hove Singularities

M. Ridene[1], T. Wassmann[2], E. Pallecchi[1], G. Rodary[1], J. C. Girard[1] and A. Ouerghi[1]

[1.] *CNRS- Laboratoire de Photonique et de Nanostructures (LPN), Route de Nozay, 91460 Marcoussis, France*
[2.] *Université Pierre et Marie Curie (CNRS-IMPMC), 4 Pl. Jussieu, 75005 Paris, France*

We report scanning tunneling microscopy and spectroscopy investigation of graphene nanoribbons grown on an array of bunched steps of a 6H-SiC(0001) substrate. Our STM images of a GNR on a step terrace feature a ($\sqrt{3}\times\sqrt{3}$)R30° pattern of aromatic rings which define our armchair nanoribbons. This is in agreement to a simulation based on density functional theory. As another signature of the one-dimensional electronic structure, in the corresponding STS spectra we find well developed, sharp Van Hove singularities.

In the ongoing efforts to investigate the properties of graphene and related structures, graphene nanoribbons (GNRs) have attracted considerable attention. To a big part, this is due to the opening of an electronic band gap caused by the confinement of the electronic states in one dimension. The boundary conditions imposed by the edges, however, give rise to more phenomena than that. In this article, we address two of them, the formation of aromatic rings of delocalized π-electrons in the real space electron distribution and the appearance of Van Hove singularities (VHSs) in the local density of states (LDOS).

Composed of a network of $sp^2$ hybridized carbon atoms, graphene can be thought of as a prototypic aromatic crystal. While three of the four valence electrons of each carbon atom are engaged in localized σ-bonds, the remaining one can form a π-bond with any of the three neighboring atoms. As long as nothing breaks the symmetry of the graphene crystal, all three options for the π-bond co-exist at the same time. When confined in narrow ribbons, however, considerations based on Clar's theory of the aromatic sextet have shown that certain π-bond configurations are preferred over others [1, 2]. Depending on the edge configuration and the width of the ribbon, a distinct ($\sqrt{3}\times\sqrt{3}$)R30° pattern of aromatic rings can form. As the states of the π-electrons lie close to the Fermi energy, this superstructure is detectable in STM measurements and has been observed near graphene edges [3]. Another effect of the one dimensional confinement of the electrons in GNRs is the appearance of divergences in the density of states (DOS) as a function of the energy, the so-called VHSs. They appear when a new sub-band in the electronic structure enters the bias window. Even though predicted theoretically, VHSs remain difficult to measure and when they are observed they are often suppressed by edge disorder [4, 5].

Contrary to other graphene growth methods, graphitization of SiC does not require a graphene transfer since SiC represents itself a suitable substrate for high power, high frequency, and opto-electronic devices [6]. The structure and electronic properties of graphene grown on SiC, however, strongly depend on the presence of steps on the substrate [6]. These steps, concentrated or bunched in narrow regions separated by larger atomically flat (0001) terraces, are known to be detrimental for the mobility of graphene. On the other hand, since the widths of the terraces are homogeneous to a high degree [7], they have proven to be useful for the patterning of nanostructures such as GNRs [7, 8].

In this study, we took advantage of such a group of bunched steps on the SiC substrate to epitaxially grow narrow, highly crystalline GNRs. Low temperature Scanning Tunneling Microscopy/Spectroscopy (STM/STS) were used to study the structure and the electronic properties of the GNRs grown on a one-dimensional array of bunched steps of a 6H-SiC(0001) substrate. At the one-dimensional array of bunched steps, we find that epitaxial GNRs are naturally formed on the few nanometers narrow terrace separating two consecutive step edges of the SiC. The epitaxial nature of these GNRs is reflected in the high crystallographic order of their armchair edges, which allow for the observation of the distinct (√3x√3)R30° pattern associated with aromatic rings in the GNRs and sharp, well defined VHSs.

In order to remove damages caused by the surface polishing, our n-doped 6H-SiC(0001) wafer was first etched in a $H_2$ atmosphere at a pressure of 200 mbar and a temperature of 1500°C for 15 minutes. The sample was prepared by electron bombardment heating under ultra high vaccum (≈ 2 $10^{-9}$ mbar) [7, 9]. To remove the native oxide, the wafer was annealed under Si flux at 900°C. For annealing at higher temperatures, we observed the gradual formation of (3x3), (√3x√3)R30° and (6√3x6√3)R30° surface reconstructions as precursors to the formation of the graphene layer. The wafer was cooled down to room temperature and transferred ex-situ to the STM chamber where the sample was outgassed at 600°C for one hour at a pressure of 5 $10^{-10}$ mbar. STM/STS measurements were carried out using an Omicron ultra high vacuum low temperature scanning tunnelling microscope (UHV-LT-STM). STM/STS were acquired at 4.2 K in the constant current mode for different bias voltages V applied to the sample. For the STS measurements, performed at $T$ = 4.2 K, the I(V) characteristics were acquired while the feedback loop was inactive, the differential conductivity dI/dV (V, x, y), proportional to the LDOS, was measured directly by using a lock-in technique. For this purpose a small AC modulation voltage $V_{mod}$ was added to $V$ ($V_{mod,p-p}$=10 mV, $f_{mod}$=973 Hz) and the signal $dI$ detected by the lock-in amplifier was used to determine the differential conductivity $dI$ /$dV_{mod}$. All STM/STS results presented here were analysed using the WSXM software [10]. The simulated STM images were obtained within the framework of Density Functional Theory (DFT) with the post-processing code of the Quantum-ESPRESSO package [11] which uses the Tersoff-Hamann approximation [12, 13], as described in Ref. 1. This calculation relied on plane-wave basis functions, Vanderbilt ultrasoft pseudo-potentials [14] and the Perdew-Burke-Ernerhof generalized-gradient approximation (GGA) to the exchange-correlation functional [15]. In these simulations, the distance between sample and tip was fixed at 3 Å.

Our procedure to produce epitaxial GNRs is illustrated in Fig. 1. It relies on two preconditions that the SiC substrate can fulfil. The first one is that the initial terraces on the substrate are atomically flat and separated by fairly straight steps (Fig. 1(a)). The second precondition is the ability to "bunch" or concentrate the steps in a narrow area (Fig. 1(b)). In our approach this is done by exposing the substrate to a $H_2$ atmosphere, which leads to the formation of areas with bunched steps on one hand and wider main terraces with (0001) orientation on the other hand (Fig. 1(b)). The graphene layer was then grown at a temperature of 1200°C. In this step, the control of the temperature and the duration of thermal annealing under $H_2$ atmosphere as well as the growth of graphene are essential. Even though graphene grows both on the large main terraces and on the narrow steps, the layer can be discontinuous at the steps of the substrate as we will argue later (Fig. 1(c))

Fig. 2(a) represents a large scale STM image of our sample after this process. The image shows flat terraces that are separated by straight step edges. In the area of bunched steps, highlighted in Fig. 2(a), we observe that the steps delimit large and narrow areas of graphene. The edges of the steps can be either parallel or secant. From the analysis of several images, we found that the narrow terraces in the region of bunched steps are about 6 to 10 nm wide. We also notice the presence of the C-rich (6√3x6√3)R30° (6R3) interface layer uncovered with graphene, this is not surprising since the substrate was intentionally partially graphitized. This interface layer is strongly bound to the substrate and does not share the electronic properties of graphene. In particular, this layer shows a poor conductivity due to the $sp^3$ nature of its atomic bonds [16].

Fig. 2(b) shows an array of parallel, bunched steps at higher magnification. These steps separate two large (0001) main terraces. The STM image of a large main terrace (Fig. 2(c)) reveals a regular honeycomb lattice, indicating that the substrate is indeed covered with a single layer of graphene. Superimposed on the graphene lattice, a hexagonal pattern with a periodicity of 1.8 nm appears. This moiré is characteristic of epitaxial graphene and stems from the interference between the graphene lattice and the interface layer situated below the graphene sheet.

In order to further probe the local electronic structure of the sample, we performed STS measurements [12]. The inset of Fig. 2(c) displays the corresponding spectrum recorded on the graphene layer covering the main terrace. This spectrum is characteristic for single-layer epitaxial graphene. The local minimum in the LDOS indicated by the arrow is known to be to the position of the Dirac point ($E_D$) which is shifted with respect to the Fermi energy due to the strong charge transfer between graphene and the SiC substrate [17]. We find a Dirac Energy of $E_D \approx$ -300 meV which implies a carrier concentration on the order of $10^{13}$ cm$^{-2}$, characteristic of epitaxial monolayer graphene [18]. Carrier concentrations in this order are rather high in comparison to the ones measured for graphene produced by exfoliation or chemical vapor deposition (CVD) and can be reduced for instance with molecular doping [19].

We turn now to the graphene grown on the narrow terraces of the bunched steps (Fig. 2(b)). In these areas, occasionally we observed a defect line parallel to the SiC steps appearing as a bright line in the STM image (Fig. 2(b)). In this case, the defect line runs in [1100] direction and divides the graphene layer into two structurally and electronically discontinuous portions [20]. A GNR is delimited by the defect line and a step edge. The ribbon contains approximately 25 hexagons in the transverse direction for a total width of about 6 nanometers. Thanks to the high quality of the SiC substrate and to the absence of a graphene transfer, the ribbon and its edges at the substrate step are flat. This simplifies the interpretation of the STM/STS data and the comparison with the theory. In presence of rolled edges, the situation becomes more complicated as the topography of bent edges can modify the band structure as has been observed for GNRs produced by unzipping of CNTs or grown by CVD [5, 21].

We now focus on a different region at the bunched steps, where a step separates two nanoribbons (Fig. 3(a)). While graphene can grow continuously across a step edge of the SiC substrate [22], in this case the electronic structure is clearly discontinuous. The discontinuity of the electronic states in the graphene layer is evidenced by the (√3×√3)R30° interference pattern of bright rings in proximity of the edges [3]. This (√3×√3)R30° superstructure has been reported in other experiments and always is related to armchair edges [23-25]. The (√3×√3)R30° periodicity of the interference pattern in the STM image can also be observed in the Fast Fourrier transform of the STM image (Fig. 3(b)): we identify a first group of spots corresponding to graphene reciprocal lattice (RL) (the corresponding lattice parameter is about 0.23 nm) and a second set of spots reflecting the (√3×√3)R30° superstructure. Similar interference patterns are a strong indication of a structural discontinuity of the graphene lattice at the step since it is not observed when the graphene film grows continuously across the substrate step [22-24, 26].

In Fig. 3(c) we show the STS spectra obtained at the different positions marked in Fig. 3(a). All of them exhibit a particular feature which is the presence of sharp peaks in the LDOS at well defined energies, separated by around 320 mV. These peaks, known as VHSs, are a signature of the one-dimensional electronic structure and were observed for CNTs [27, 28] and recently also for GNRs [4, 5]. The step-like increase in the LDOS for energy larger than the VHS arises from the density of states of the new one-dimensional subband entering the bias windows. Interestingly, the first pair of VHSs is not completely symmetric with respect to the Fermi energy, but an asymmetry of about 50 meV is present. This is probably due to the charge transfer from the substrate to the GNRs in a similar way to what has been observed in the case of CNTs [27, 29]. We will define the energy $E_g$ as the distance between the first two VHSs [27]. Our $E_g$ can suit with the theoretical considerations which have shown that the bandgap of GNRs is inversely proportional to their width and can be approximated by the formula $E_g(W)=\pi\hbar v_F/W \approx (2\text{ eV nm}/W)$ [4, 6], where $v_F$ is the Fermi velocity and W is the width of the ribbon. For $E_g$ =320 meV this relation gives a width of $W \approx 6$ nm, in line with our observation that the steps are typically 6 to 10 nm wide.

Finally we present a direct comparison of our measured image of the nanoribbon with the simulated STM signal of a 3.9 nm wide, hydrogen terminated armchair GNR (Fig. 4(a)). The theoretical STM image is in excellent qualitative agreement to the pattern observed in the experiment. This confirms the armchair termination of our nanoribbon. We also calculated the LDOS for the 3.9 nm wide nanoribbon (Fig. 4(b)). To account for disorder, we used a broadening of 34 meV, the theoretical LDOS is in qualitative agreement with the experimental one, with Van Hove singularities at the onsets of the subsequent energy subbands. It should be noticed that DFT-GGA calculation is known to underestimate the value band gap [1].

To summarize, we have found that the step bunching of SiC can be used as a template for the growth of well defined, narrow GNRs with highly ordered edges. The quantum interference pattern in the STM images show that the graphene layer in our sample is electronically discontinuous at the steps of the SiC substrate. The sharp VHSs and the step like increases in the LDOS confirm the one-dimensional nature of the GNRs. The STM images of the GNRs on the narrow steps feature a characteristic (√3×√3)R30° pattern of aromatic rings. This pattern can be understood in terms of Clar's theory of the aromatic sextet and is in very good agreement with our DFT calculation for an armchair nanoribbon.

We acknowledge B. Etienne and F. Mauri for fruitful discussions. This work was supported by the French Contract No. ANR-2010-BLAN-0304-01-MIGRAQUEL and the French-Tunisian CMCU project 10/G1306.

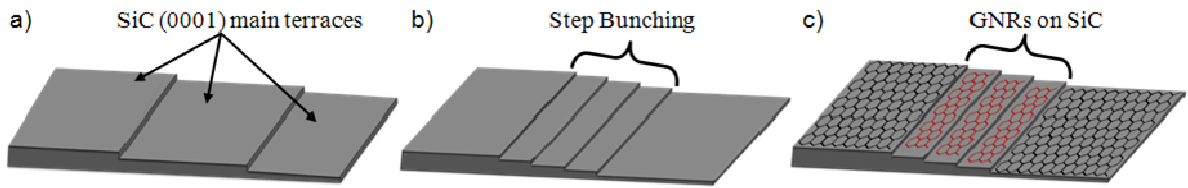

**Figure 1:** Step-by-step illustration of the synthesis of epitaxial GNRs on SiC(0001): **(a)** Pristine 6H-SiC(0001) substrate with regular steps. **(b)** Bunching of steps following the etching under hydrogen atmosphere. **(c)** Growth of GNRs on the bunched steps and of graphene on the main terraces.

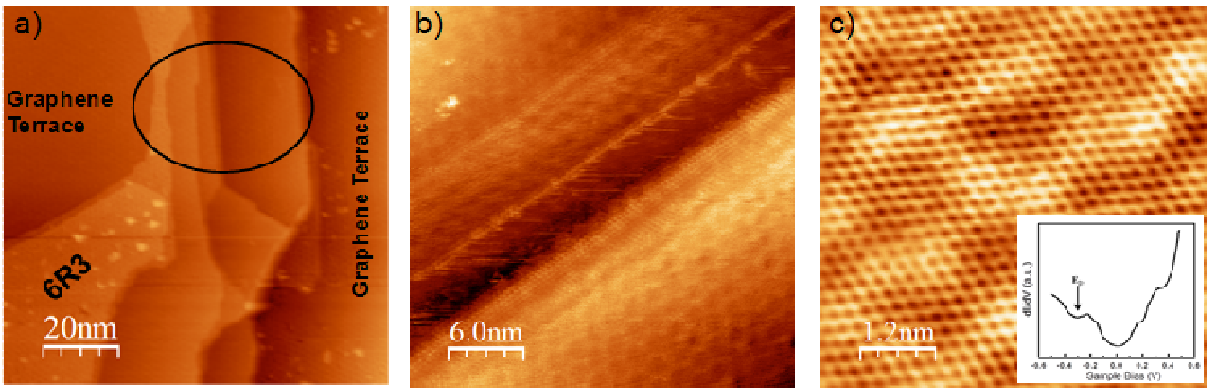

**Figure 2:** Signature of epitaxial GNRs by STM topographic images: **(a)** Large scale STM topographic image (100 x 100 nm$^2$, -2 V, 500 pA) showing an area of bunched steps. **(b)** Low temperature STM image (30 x 30 nm$^2$, 50 mV, 500 pA) in the vicinity of a step bunching. **(c)** Atomic resolution STM image of the graphene lattice taken on the main terrace. **Inset:** STS spectrum of this region (the Fermi energy is at V= 0 V) showing the typical electronic structure of epitaxial graphene on SiC.

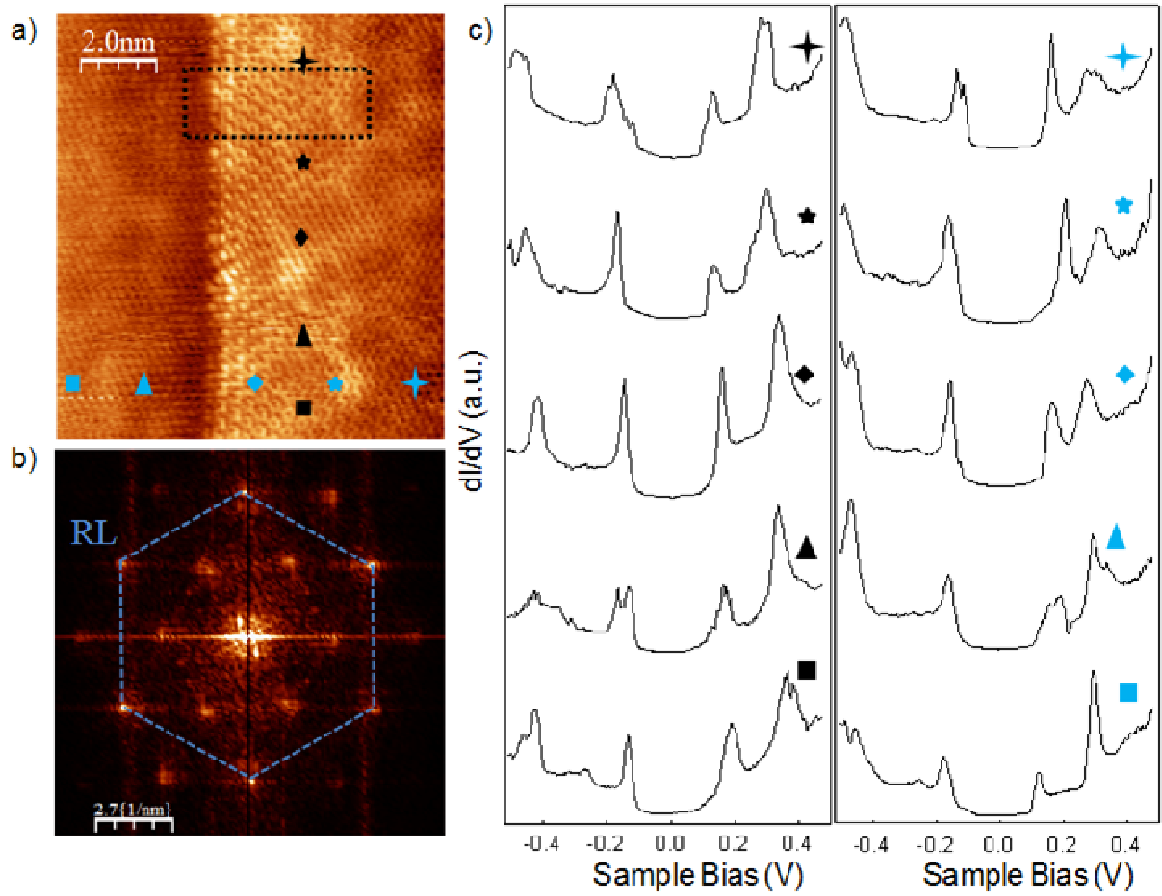

**Figure 3:** Electronic structure of epitaxial GNRs: **(a)** STM image (10 x 10 nm², 50mV, 500 pA) at a step edge. **(b)** Fast Fourrier Transformation of **(a)** showing the periodicity of the graphene lattice as well as the one of the (√3×√3)R30° pattern induced by the edge. **(c)** STS spectra taken at the positions marked in **(a)**. Sharp VHSs observed all over the spectra revealing a one-dimensional electronic structure.

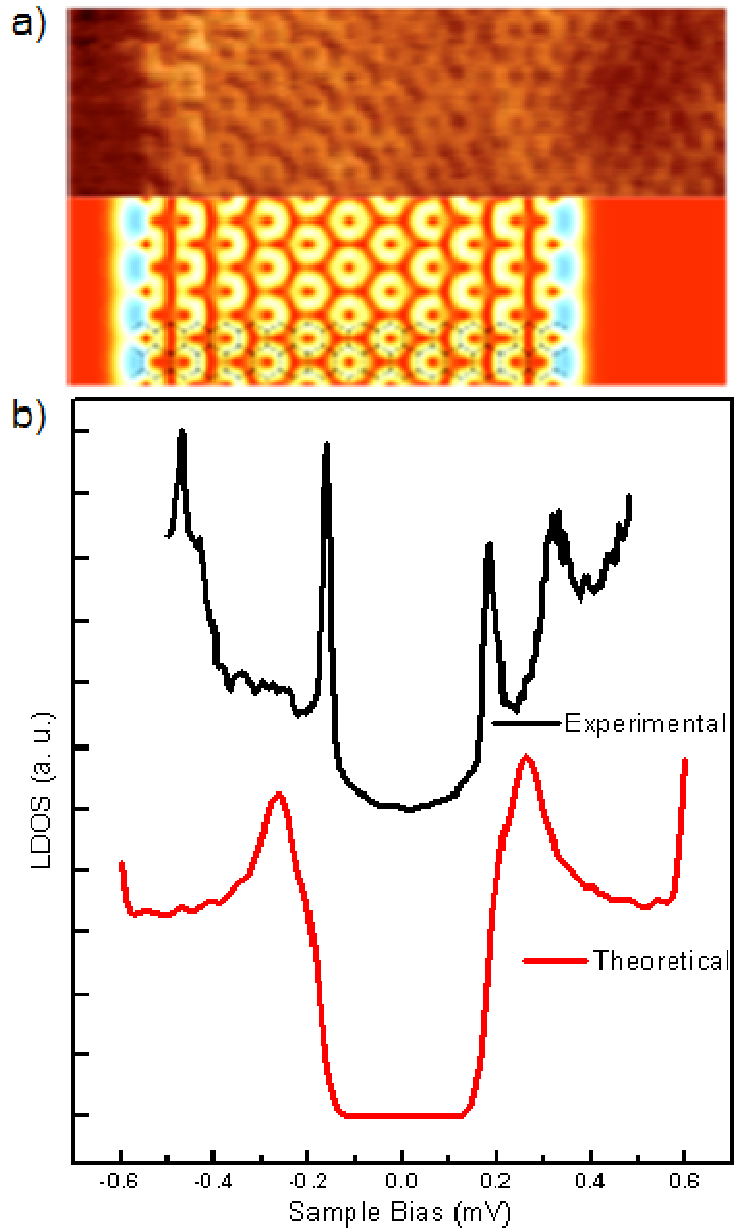

**Figure 4: (a)** Side-by-side comparison of the area indicated by the dashed frame in Fig 3**(a)** with a simulated STM image (bias 350 mV) of a hydrogen terminated armchair GNR. The excellent agreement between the two images is an indication for the good quality of the epitaxial GNR edge. **(b)** Experimental vs Theoretical STS spectra